\documentclass{llncs}

\title{{\bf Part of Speech Based Term Weighting for Information Retrieval}}
\titlerunning{}

\author{Christina Lioma \inst{1}
	\and 
	Roi Blanco \inst{2}}
\authorrunning{Christina Lioma \and Roi Blanco}
\tocauthor{Christina Lioma (Katholieke Universiteit Leuven),
Roi Blanco  (La Coruna University)}

\institute{Computer Science, Katholieke Universiteit Leuven, 3000, Belgium
\and Computer Science, La Coruna University, 15071, Spain
\email{christina.lioma@cs.kuleuven.be, rblanco@udc.es}}

\newcommand\blfootnote[1]{%
  \begingroup
  \renewcommand\thefootnote{}\footnote{#1}%
  \addtocounter{footnote}{-1}%
  \endgroup
}

\usepackage{ulem} %allows me strike through text (with \sout{})
\usepackage{color}
\usepackage{colortbl}
\usepackage{graphics}
\usepackage{graphicx}
\usepackage{multirow}
\usepackage{latexsym}
\usepackage{amsmath}
\usepackage{lscape}
\usepackage{hyperref}

\begin{document}

\maketitle
\begin{abstract}
Automatic language processing tools typically assign to terms so-called `weights' corresponding to the contribution of terms to information content. 
Traditionally, term weights are computed from lexical statistics, e.g., term frequencies. 
We propose a new type of term weight that is computed from part of speech (POS) n-gram statistics.
The proposed POS-based term weight represents how informative a term is in general, based on the `POS contexts' in which it generally occurs in language. We suggest five different computations of POS-based term weights by extending existing statistical approximations of term information measures. We apply these POS-based term weights to information retrieval, by integrating them into the model that matches documents to queries. 
Experiments with two TREC collections and 300 queries, using TF-IDF \& BM25 as baselines, show that integrating
our POS-based term weights 
to retrieval always leads to gains (up to +33.7\% from the baseline). 
Additional experiments with a different retrieval model as baseline (Language Model with Dirichlet priors smoothing) and our best performing POS-based term weight, show retrieval gains always and consistently across the whole smoothing range of the baseline. 
\end{abstract}

\section{Introduction}
\blfootnote{Preprint of: Christina Lioma and Roi Blanco. Part of Speech Based Term Weighting for Information Retrieval. In: Advances in Information Retrieval, 31th European Conference on {IR}
               Research, {ECIR} 2009, Toulouse, France, April 6-9, 2009. Proceedings. Ed. by Mohand Boughanem and
               Catherine Berrut and
               Josiane Mothe and
               Chantal Soul{\'{e}}{-}Dupuy. Springer, 2009, pp. 412--423. isbn: 978-3-642-00957-0. doi: \url{10.1007/978-3-642-00958-7_37}.
}

With the increase in available online data, accessing relevant information becomes more difficult and provides a strong impetus for the development of automatic language processing systems, able to convert human language into representations that can be processed by computers. %, with the aim to provide humans with access to  knowledge. 
Typically, these systems locate and quantify information in data by making statistical decisions about the occurrence and distribution of words in text. These statistical decisions have led to the development of term weights which reflect how informative a word is within some text, e.g., the well-known Inverse Document Frequency (IDF) weight~\cite{KSJ:1972}.

	We propose an alternative type of term weight, computed from part of speech (POS) information (e.g., verb, noun), and specifically POS n-grams. 
	These POS-based term weights represent how informative a term is in general, based on the `POS contexts' in which the term occurs in language.
	The motivation for using POS is that their shallow grammatical information can indicate to an extent the presence or absence of content. This is a well-known grammatical notion, for instance Jespersen's Rank Theory uses this notion to semantically define and rank POS~\cite{Jespersen:1929}.  
	The motivation for using n-grams is their well-known language modelling advantages: representing `small contexts' (inside the n-gram), and profiling `large samples' (the collections from which they are extracted).
	The intuition behind our POS n-gram based term weights is to reward terms occurring often in content-rich POS n-grams, which are n-grams of salient POS, such as nouns, verbs or adjectives. %to compute a term weight.  

We apply our POS-based term weights to Information Retrieval (IR), by integrating them into the retrieval model that matches documents to user queries, using a standard way of integrating additional evidence into retrieval \cite{CraswellR:2005}. Using the original retrieval model as a baseline, and experimenting with three established models (TF-IDF, BM25, LM with Dirichlet priors smoothing), two standard TREC~\cite{VoorheesH:2005} collections, and 300 queries, we see that integrating our POS-based term weights to retrieval enhances performance notably, with respect to average and early precision, and with a statistical significance at most times.
 
 The contributions of this work are: (i) it proposes a type of term weight that is derived from POS n-grams, (ii) it shows that this POS-based term weight can be integrated to IR, similarly to additional evidence, with benefits to retrieval performance. 
  
This paper is organised as follows.
	Section~\ref{s:POS} discusses the motivation for computing a term weight from POS.  
	Section~\ref{s:Relstud} presents other applications of POS n-grams and related work on term weighting.
	Section~\ref{s:PIS} describes how we derive a term weight from POS n-grams, and 
	Section~\ref{s:Evaluation} evaluates our proposed term weights in the IR task. 
	Section~\ref{s:Conclusion} concludes this work.

\section{Motivation for using parts of speech}
\label{s:POS}
Our motivation is that the shallow grammatical information carried by POS can indicate to an extent the presence or absence of informative content. 
This is certainly not new; it can be found as an observation in 4$^{th}$ century BC studies of Sanskrit~\cite{Lyons:1977}, and also formalised into a linguistic theory for ranking POS \cite{Jespersen:1929}. 
In fact, Jespersen's Rank Theory suggests that POS are semantically definable and subject to ranking according to \textit{degrees} \cite{Jespersen:1929}: 
firstly (most content-bearing) nouns; secondly adjectives, verbs and participles; thirdly adverbs; and finally all remaining POS.
Jespersen's notion of degree is defined in terms of the combinatorial properties of POS: each POS is modified by a higher degree POS, e.g., nouns are modified by verbs, and verbs are modified by adverbs. %No more than three degrees were required, because no major POS modifies a third degree POS. 

A more general POS distinction is between major (or open) and minor (or closed) POS, where roughly speaking open POS mainly bear content, and closed POS mainly modify content. Open POS correspond to Jespersen's first, second, and third degrees, and closed POS correspond to the remaining POS. This POS distinction is not language-dependent e.g., Chinese grammatical theory also traditionally distinguishes between `full' and `empty' words~\cite{Lyons:1977}. In addition, this distinction can have philosophical  extensions, for instance, it can be compared to the Aristotelian opposition of `matter' and `form', with open  POS signifying objects of thought which constitute the `matter' of discourse, and closed POS contributing to the meaning of sentences by imposing upon them a certain `form' or organisation~\cite{BasD:2004}. A more practical implementation of this distinction is language processing systems that consider closed POS as `stopwords' and exclude them from processing.

In this light, POS n-grams can become `POS contexts' for which we have some prior knowledge of content, e.g. POS n-grams containing nouns and verbs are likely to be more informative than POS n-grams containing prepositions and adverbs. 
We look at all the POS n-grams of a term and we reason that the more informative and frequent these POS n-grams are, the more informative that term is likely to be.
We propose to use such a general `term informativeness' term weight in IR, motivated by the fact that similar notions are often used in readability formulae to predict the comprehension or complexity level of texts~\cite{Mikk:2001}. %For instance, word commonness is used in theoretical and practical linguistics, e.g. in quantitative linguistics, lexicography, and language teaching~\cite{SavickyH:2002}. %In fact, \cite{Mikk:2001} suggests a `corrected term frequency' based on word commonness. 
%\begin{table} 
%\begin{center} 
%\resizebox{120mm}{!}{
%\begin{tabular}{l} 
%MAJOR POS: adjective (JJ), noun (NN), participle (VR), verb (VB)\\
%\hline
%MINOR POS: adverb (RB), conjunction (CC), determiner (DT), modal verb (MD), numeral (CD),\\
%particle (RP), possessive ending (POS), preposition (IN), pronoun (PP), symbol (SY)\\ 
%\end{tabular}
%} 
%\end{center} 
%\caption{
%	\label{tab:treebank} Penn TreeBank: primary parts of speech.} 
%\end{table}

\section{Related work}
\label{s:Relstud}	
\subsection{Applications of part of speech n-grams}
POS have been used for a variety of different applications. 
In POS tagging, they are used to predict the POS of a word on the basis of its immediate context, modelled within the n-gram~\cite{BrownP:1992}. Several well-known POS taggers use POS n-grams, e.g. Mxpost \cite{Ratnaparkhi:1996} or TreeTagger \cite{Schmid:1997}. Another application is stylometric text categorisation, where POS n-grams assist in predicting the author/genre of a given text. In such applications, POS n-grams, described as `pseudo-word sequences' \cite{BaayenH:1996}, `POS triplets' \cite{SantiniP:2006}, or `quasi-syntactic features' \cite{KoppelA:2003} have been used with promising results. 
POS n-grams are also used in IR, for instance to prune term n-grams from IR system indices in order to reduce storage costs \cite{LiomaO:2007b}, or to predict the difficulty of search terms \cite{AslamP:2007}. 
In machine translation, POS n-grams are often used to select the best translation among several candidates \cite{HwaR:2002}, for instance by looking at the more likely correspondence of POS patterns between the source and target languages.
Spell or grammar checking \cite{WagnerF:2007} and automatic summarisation \cite{CorstonR:2004} also use POS n-grams.   

Overall, POS n-grams are typically used to predict the occurrence of an item in a sequence (e.g., POS tagging), or to characterise the sample from which they are extracted (e.g., text classification). Using POS n-grams to compute a term weight differs from that. Recently, Lioma \& van Rijsbergen \cite{LiomavR:2008} proposed deriving term weights from POS n-grams, using Jespersen's Rank Theory. 
Specifically, they adapted Jespersen's POS ranks into POS weights, the values of which were tuned empirically. 
In this work, we present five different POS-based weights. Our weights are computed from POS n-grams, and in this respect they are similar to the work proposed in~\cite{LiomavR:2008}. However, whereas in \cite{LiomavR:2008} term weights are derived according to Jespersen's theory for ranking POS, we derive all five proposed weights by extracting POS n-gram statistics directly from the collection.
Hence, our approach is not based on a linguistic theory like in \cite{LiomavR:2008}, but on collection statistics.
Practically this means that whereas in \cite{LiomavR:2008} Lioma \& van Rijsbergen employ four different parameters, which they tune in order to optimise retrieval performance, our proposed weights are parameter-free in this respect. In Section~\ref{ss:ExperimentalResults} we present experiments using both the POS weights proposed by \cite{LiomavR:2008} and the POS weights proposed in this work, and we show that the latter outperform the former. 

\subsection{Term weighting schemes}
\label{ss:termweighting}
Typically, term weighting schemes assign to terms weights which represent their contribution to the contents of some document or collection of documents. 
A popular term weight is IDF~\cite{KSJ:1972}, which weights how discriminative a term is by looking at how many different documents contain it in a general collection of documents:
$idf = \log{\frac{N}{df}}$, where $N$ is the total number of documents in the collection, and $df$ is the number of documents that contain a term (\textit{document frequency}).
The intuition behind IDF is that the more rare a word is, the greater the
chance it is relevant to those documents in which it appears.

Other term weights have also been proposed. 
Bookstein \& Swanson introduced the $x_I$ measure for a word $w$ \cite{BooksteinS:1974}:
$x_I (w) = tf - df$, where $tf$ is the frequency of a word in a document. 
This term weight is intuitive to an extent (e.g., for two words of the same frequency, the more
`concentrated' word will score higher), but can be biased toward frequent words, which tend to be less informative~\cite{Papineni:2001}.

In \cite{Harter:1975} Harter proposed another term weight, called z-measure in an earlier formulation by Brookes~\cite{Brookes:1968}, based on the observation that informative
words tend to divert from a Poisson distribution. He suggested that informative words may be identified by observing their fit to a mixture of a 2-Poisson model. The z-measure computes the difference between the means of the two distributions, divided by the square-root of the respective summed variances.
Harter found this term weight successful for keyword identification in indexing. Eventually, this approach was %superseded 
 extended by N-Poisson mixtures~\cite{Margulis:1992}.

More recently, Cooper et al. suggested an extension of IDF \cite{CooperC:1993}: they used logistic regression to assign term weights according to how often a
query term occurs in the query and a document, the term's IDF,  and the number of distinct terms common to both query
and document. 
Another IDF extension was suggested by Church \& Gale \cite{ChurchG:1995}. Their alternative, \textit{Residual IDF} (RIDF), was motivated by the observation that nearly
all words have IDF scores that are larger than what one
would expect according to an independence-based model
(such as Poisson):
$RIDF = IDF - \widehat{IDF}$, where $\widehat{IDF}$ is the expected IDF.
Rennie \& Jaakkola note that even though the RIDF intuition is similar to that of the $x_I$ measure, $x_I$ has a bias toward high-frequency words,
whereas RIDF has the potential to be largest for medium-frequency words, and as such may be more effective \cite{RennieJ:2005}.

Over the years several variations of term frequency heuristics have been used for term weighting. Some of those are
the lnc.ltc weight by Buckley et al.~\cite{BuckleyS:1995}; the approach of Pasca to distinguish
between terms of high, medium and low relevance using heuristical rules \cite{Pasca:2001};
the extension of Pasca's heuristics by Monz, who used machine learning techniques to learn term
weights by representing terms as sets of features, and applied the resulting term weights to question answering \cite{Monz:2007}.

All these approaches have one overriding factor in common: they attempt to capitalise on the frequency and distribution of individual terms in the collection in order to provide a statistical estimate of the importance of a term in a document/collection, aside of the semantic or syntactic nature of the term itself.
The POS-based term weights we propose are different in this sense, because the usage of the term is captured and considered in order to determine the term's importance, as opposed to only considering occurrence data, and variations of.

\section{POS-based term weighting}
\label{s:PIS}
The aim is to suggest term weights, which represent how informative a term is, and which have been computed from POS n-gram information only. 
The general methodology is: (1) `Map' terms to the POS n-grams that `contain' them\footnote{A POS n-gram that `contains' a term = a POS n-gram which corresponds to a term n-gram that contains a term.} and store their frequency statistics.  
(2) Approximate the probability that a term is informative on the basis of how informative its corresponding POS n-grams are. 

Let $\{pos\}$ be the set of parts of speech, and $\{POS\}$ the set of POS n-grams, where if $POS \in \{POS\}, POS=[pos_1,\dots pos_N], pos_i \in \{pos\} \forall i \in [1 \dots N] $. Let $I$ be a random variable for informative content. % ($I$ can take values 0 or 1, just like relevance in IR).
Also, let $\{POS\}_t$ be the set (no duplicates) of POS n-grams that `contain' term $t$. Then, according to the total probability theorem, the probability that a term is informative $P(I|t)$ is:
\begin{align}
\label{eq:posframe1}
p(I|t) &= \sum_{POS \in \{POS\}} p(I|t,POS)\; p(POS | t) \approx \sum_{POS \in \{POS\}_t} p(I|t,POS)\; p(POS|t)%\\ 
\end{align}
\noindent where we assume that $p(POS | t) = 0$ if $POS \notin \{POS\}_t$. Otherwise, there are two options to compute $p(POS | t)$:
(i) the probability can be considered uniform, regardless of how many times a POS n-gram `contains' the term (boolean option);
(ii) the probability can be estimated by counting POS n-gram frequencies in the collection (weighted option).

The five different weights we propose have Eq.~\ref{eq:posframe1} as a starting point, but compute its components in different ways. We present these weights in Sections~\ref{ss:ml}-\ref{ss:heur}, and show how we integrate then into the retrieval model using a standard formulation for integrating evidence into retrieval~\cite{DBLP:conf/sigir/TaoZ07} in Section~\ref{ss:integration}. 

\subsection{POS n-gram Maximum Likelihood}
\label{ss:ml}
We derive a term weight using Eq.~\ref{eq:posframe1}, and approximate $p(I|t,POS) \approx p(I|POS)$ by computing the maximum likelihood (ML) of individual POS n-grams in a collection. This is similar to building a language model of POS n-grams from their occurrence in a collection $\mathcal{C}$:
$p(I|POS) \propto p(POS | \mathcal{C})$. This equation assumes that the informative content of a POS n-gram is approximately proportional to the frequency of a POS n-gram in a collection. This approximation is parameter-free.
The above produces two different weights, when combined with a boolean and respectively weighted option for computing $p(POS | t)$. We call these weights \textbf{pos\_ml\_boolean} and \textbf{pos\_ml\_weighted} respectively. 

\subsection{POS n-gram Inverse Document Frequency (IDF)}
\label{ss:heur}
We also suggest three alternative term weight computations using POS n-gram statistics, inspired by the computations of IDF, RIDF and Bookstein and Swanson's $x_I$, presented in Section~\ref{ss:termweighting}.

\noindent \textbf{pos\_idf:}
In conventional term IDF, \textit{document frequency} (df) is the number of documents in which a term occurs. In our proposed pos\_idf, we count the number of POS n-grams in which a term occurs, and refer to this as \textit{POS ngram frequency} (pf). 
We compute pos\_idf as follows: $pos\_idf = log \frac{|C|}{pf}$,
where $|C|$ is the number of all POS n-grams in the collection.
Note that pos\_idf can be effectively derived from Eq.~\ref{eq:posframe1} if we consider
 $p(POS | t) = 1/\{POS\}_t$ and $p(I|t,POS)=1$ if $POS \notin \{POS\}_t$ and 0 otherwise.

\noindent \textbf{pos\_ridf:}
We compute the Residual IDF (RIDF) of POS n-grams: $pos\_ridf = pos\_idf - \widehat{pos\_idf}$,
where pos\_idf is computed with the equation shown immediately above, and 
$\widehat{pos\_idf}$ is the expected $pos\_idf$, computed as $-log(1 - e^{TF/|C|})$, where $TF$ is the number of times a term occurs in the different POS n-grams, and $|C|$ is as defined above.

\noindent \textbf{pos\_bs:}
We compute Bookstein and Swanson's term weight of POS n-grams:
$pos\_bs = TF - pf$, where 
$TF$ and $pf$ are as defined above.
In this paper we take the log of $TF - pf$ to compute our term weight.

\subsection{Integration into retrieval models}
\label{ss:integration}
In Sections \ref{ss:ml}-\ref{ss:heur}, we suggest in total five POS-based term weights, namely:
\begin{enumerate}
\item \textbf{\small pos\_ml\_boolean:}	{\small POS n-gram Maximum Likelihood; ignores how often a POS n-gram `contains' a term}
\item \textbf{\small pos\_ml\_weighted:} {\small POS n-gram Maximum Likelihood; considers how often a POS n-gram `contains' a term}
\item \textbf{\small pos\_idf:} {\small how many POS n-grams `contain' a term}
\item \textbf{\small pos\_ridf:} {\small how many POS n-grams `contain' a term}
\item \textbf{\small pos\_bs:} {\small term frequency in POS n-grams, how many POS n-grams `contain' a term}
\end{enumerate} 

Our POS-based term weights are document-independent weights that measure the general (non-topical) informative content of terms. 
We integrate them into the retrieval model, using a standard integration of document-independent evidence into retrieval, \cite{CraswellR:2005} or term proximity evidence \cite{DBLP:conf/sigir/TaoZ07}: 
\begin{equation}
New\_score(t,d) = Old\_score(t,d) + w \cdot pos\_weight
\label{eq:craswell}
\end{equation}
\noindent where $New\_score(t,d)$ (resp. $Old\_score(t,d)$) is the score of a document for a query that integrates (resp. does not integrate) our POS-based term weight, $w$ is a parameter that controls the integration, and $pos\_weight$ is our POS-based term weight.
When combining evidence in this way, we combine evidence that is dependent on the query (as in \cite{DBLP:conf/sigir/TaoZ07}), and not query independent evidence (as in \cite{CraswellR:2005}). Note that the type of evidence, query independent or not, is arbitrary. In \cite{CraswellR:2005} Craswell et al. proposed various ways of integrating evidence into retrieval. Here, we employ their simplest way which contains one parameter only (extended from Eq. 1 in \cite{CraswellR:2005}).
Other ways of integrating our POS-based term weights to retrieval are also possible, e.g., by rank merging, or as prior probabilities~\cite{CraswellR:2005}.

\section{Evaluation}
\label{s:Evaluation}
\textbf{Experimental methodology:} 
We integrate the POS-based term weights into retrieval models, and compare retrieval performance against a baseline of the retrieval models without our POS-based weights. %In addition, we present results with the POS-based weight proposed by Lioma \& van Rijsbergen in \cite{LiomavR:2008}. Note that when doing so, we integrate that POS-based weight using equation \ref{eq:craswell}, and not the integration originally presented in \cite{LiomavR:2008}.
In addition, we present results with the POS-based weight proposed by Lioma \& van Rijsbergen in \cite{LiomavR:2008}, so that we can compare directly the POS weight of \cite{LiomavR:2008}, which is derived from a linguistic theory, to our proposed POS weights, which are derived from collection statistics. Note that when doing so, we integrate the POS-based weight of \cite{LiomavR:2008} using Eq.~\ref{eq:craswell}, and not the integration originally presented in \cite{LiomavR:2008}.

We conduct three rounds of experiments: (1) We integrate our five proposed POS-based term weights to TF-IDF \& BM25, and we tune the parameter $w$ of the integration separately for each POS-based term weight (x5), retrieval model (x2), collection (x2), and evaluation measure (x2). We tune $w$ by ranging its values between [0-50000]. 
(2) We further test the robustness of our POS-based weights as follows. For Disks4\&5 only (the collection with the most queries), we train our POS-based weights on 150 queries (301-450), and test on the remaining 100 queries (601-700). Each time we train by tuning parameter $w$ of the integration separately for MAP and P@10.
(3) We further experiment with an additional baseline model (LM with Dirichlet priors smoothing) and our best performing POS-based term weight (pos\_ml\_weighted). Here we tune both the parameter $w$ of the integration and the smoothing parameter $\mu$ of the LM.
\begin{table}
\caption{Collection features.}
\begin{center}
\resizebox{75mm}{!}{
\begin{tabular}{|r|r|r|r|r|r|}
\hline
collection	&domain	&size		&documents	&terms (unique)	&POS 4-grams\\
\hline
Disks 4\&5	&news		&1.9GB	&528,155	&840,536	&25,475\\
WT2G		&Web		&2GB		&247,491	&1,159,310	&25,915\\
\hline
\end{tabular}}
\label{tab:collections}
\end{center}
\end{table}  

\noindent \textbf{Retrieval settings:} For retrieval we use the Terrier\footnote{ir.dcs.gla.ac.uk/terrier/} IR system, and we extend its indexing functionalities to accommodate POS n-gram indexing. We match documents to queries with three established and statistically different retrieval models: 
(1) the traditional TF-IDF~\cite{RobertsonKSJ:1976} with pivoted document length normalisation~\cite{SinghalB:1996}. We use TF-IDF with pivoted document length normalisation over standard TF-IDF because it does not include an explicit document length normalisation parameter;  
(2) the established Okapi's Best Match 25 (BM25)~\cite{RobertsonW:1994};
(3) the more recent Language Model (LM) with Dirichlet priors smoothing~\cite{CroftL:2003}.  
BM25 includes three tunable parameters: $k_1$ \& $k_3$, which have little effect on retrieval performance, and $b$, which normalises the relevance score of a document for a query across document lengths. 
We use default values of all BM25 parameters: \( k1 = 1.2 \), \( k3 = 1000 \), and \( b = 0.75 \)~\cite{RobertsonW:1994}. 
We use default values, instead of tuning these parameters, because our focus is to test our hypothesis, 
and not to optimise retrieval performance. If these parameters are optimised, retrieval performance may be further improved. 
LM Dirichlet includes a smoothing parameter $\mu$, which we tune to optimise retrieval performance (for the second round of experiments only).
Table~\ref{tab:collections} presents the two TREC~\cite{VoorheesH:2005} collections used:  Disks~4\&5 and WT2G. 
Disks 4\&5 contain news releases from printed media; this collection is mostly homogeneous (it contains documents from a single source).
WT2G consists of crawled pages from the Web, which is itself a 
heterogeneous source. %These collections also differ in their word statistics, since larger collections do not necessarily contain more unique terms. 
Even though the collections are of similar size (1.9GB - 2GB), they differ in word statistics (Disks 4\&5 have almost twice as many documents as WT2G, but notably less unique terms than WT2G), and domain (newswire, Web).
For each collection, we use its associated set of queries: 301-450 \& 601-700 for Disks 4\&5,  and 451-500 for WT2G. 
 We experiment with short queries (title) only, because they are more representative of real user queries on the Web~\cite{OzmutluS:2004}. We evaluate retrieval performance in terms of Mean Average Precision (MAP) and Precision at 10 (P10) and report the results of statistical significance testing using the Wilcoxon matched-pairs signed-ranks test. 

\noindent \textbf{POS-based term weighting settings:} We POS tag the collections with the freely available TreeTagger~\cite{Schmid:1997}. We collapse the Penn TreeBank tags used by the TreeTagger into the fourteen POS categories used in~\cite{LiomavR:2008}, because we are not interested in morphological or other secondary grammatical distinctions, but in primary grammatical units. We extract POS n-grams, and set n=4 following~\cite{LiomavR:2008}. Varying n=[3,6] is expected to give similar results~\cite{LiomavR:2008}. 

\subsection{Experimental results}
\label{ss:ExperimentalResults}
Table~\ref{table:TF-IDF} shows the retrieval performance of our experiments (best scores are bold). At all times, all five POS-based term weights enhance retrieval. This improvement is more for MAP than for P@10 (this is common in IR, because it is hard to alter the top ranks of relevant documents).  This improvement is also more notable for WT2G than for Disks 4\&5, even though we have more queries for the latter. A possible reason for this could be that WT2G contains more noise (being a Web crawl), and hence there is more room for improvement there, than in a cleaner collection like Disks 4\&5. In fact, a noisy collection is a good environment for illustrating the use of  a general informativeness term weight. The best performing POS-based term weight is using the maximum likelihood of POS n-grams in a collection (pos\_ml\_weighted). The particularly high parameter $w$ values of this weight are not indicative of any special treatment (identical tuning has been applied to all weights), but simply caused because this computation originally gave low magnitude weights.
\begin{table*}
\caption{Retrieval performance for MAP and P@10.  * (**) denote statistical (very strong) significance with Wilcoxon's $p<0.05$ (0.01). $\dagger$ denotes the POS weight proposed in \cite{LiomavR:2008}, included here for comparison. $w$ is the integration parameter.}
\centering
\label{table:TF-IDF}
\resizebox{130mm}{!}{
\begin{tabular}{|l||ll|ll||ll|ll|} 
\hline
				&\multicolumn{4}{|c|}{Disks 4\&5} 		&\multicolumn{4}{|c|}{WT2G}\\
\hline
\hline
 model				& MAP					&$w$	&P@10					&$w$	&MAP					&$w$	&P@10					&$w$ \\
\hline
TFIDF baseline 			&0.1935 				&-	&0.3855					&-  	&0.1933 				&- 	&0.3940 				&-\\
\hline
TFIDF\_pos\_jes$\dagger$ 	&0.2132** {\scriptsize (+10.2\%)}	&10	&0.4000* {\scriptsize (+3.8\%)}		&10	&0.2389** {\scriptsize(+23.6\%)}	&5K	&0.4080* {\scriptsize(+3.6\%)}		&2K \\
\hline
TFIDF\_pos\_ml\_wei 		&\bf 0.2256** {\scriptsize(+16.6\%)} 	&25K	&\bf 0.4044** {\scriptsize (+4.9\%)}	&22K	&0.2345** {\scriptsize(+21.37\%)}	&5 	&0.4020 {\scriptsize(+2.0\%)}		&1.5 \\
 TFIDF\_pos\_ml\_boo 		&0.2066** {\scriptsize(+6.8\%)} 	&1K	&0.3980* {\scriptsize (+3.2\%)}		&1K	&0.2068* {\scriptsize(+7.0\%)}		&2 	&0.3992 {\scriptsize(+1.3\%)}		&2 \\
TFIDF\_pos\_idf			&0.2190** {\scriptsize(+13.2\%)}	&3 	&0.4036** {\scriptsize (+4.7\%)}	&2  	&\bf 0.2584** {\scriptsize(+33.7\%)}	&5 	&\bf 0.4160** {\scriptsize(+5.6\%)}	&2 \\
TFIDF\_pos\_ridf		&0.2039** {\scriptsize(+5.4\%)} 	&3  	& 0.3948 {\scriptsize (+2.4\%)}		&3 	&0.2515** {\scriptsize(+30.1\%)}	&20	&0.4140* {\scriptsize(+5.1\%)}	 	&15 \\
TFIDF\_pos\_bs 			&0.2068** {\scriptsize(+6.9\%)} 	&2  	&0.3992 {\scriptsize (+3.6\%)}		&2  	&0.2479** {\scriptsize(+28.2\%)}	&15	&0.4080* {\scriptsize(+3.6\%)}		&30 \\
\hline
\hline
BM25 baseline 			&0.2146 				&-	&0.3960					&-  	&0.2406 				&- 	&0.4280 				&-\\
\hline
BM25\_pos\_jes$\dagger$ 	&0.2202** {\scriptsize(+2.6\%)}		&2	&0.4008 {\scriptsize (+1.2\%)}		&0.8 	&0.2755** {\scriptsize(+14.5\%)}	&5 	&\bf{0.4440}* {\scriptsize(+3.7\%)}	&2 \\
\hline
BM25\_pos\_ml\_wei 		&\bf 0.2267** {\scriptsize(+5.6\%)} 	&6K	&\bf 0.4016 {\scriptsize (+1.4\%)}	&3K   	&0.2710** {\scriptsize(+12.6\%)}	&10K 	&0.4300 {\scriptsize(+0.5\%)}		&200 \\
BM25\_pos\_ml\_boo 		&0.2187* {\scriptsize(+1.9\%)} 		&200	&0.4008 {\scriptsize (+1.2\%)}		&100  	&0.2661** {\scriptsize(+10.6\%)}	&1K 	&0.4380* {\scriptsize(+2.3\%)}		&200 \\
BM25\_pos\_idf			&0.2223** {\scriptsize(+3.6\%)}		&0.7 	&0.4000 {\scriptsize (+1.0\%)}		&0.2  	&\bf 0.2775** {\scriptsize(+15.3\%)}	&1.5 	&0.4380 {\scriptsize(+2.3\%)}		&0.7 \\
BM25\_pos\_ridf			&0.2163 {\scriptsize(+0.8\%)} 		&0.4	&0.3972 {\scriptsize (+0.3\%)}		&0.7 	&0.2679 {\scriptsize(+11.3\%)}		&1 	&0.4300	{\scriptsize(+0.5\%)}	 	&1 \\
BM25\_pos\_bs 			&0.2165 {\scriptsize(+0.9\%)} 		&0.1	&0.3964 {\scriptsize (+0.1\%)}		&0.1  	&0.2693 {\scriptsize(+11.9\%)}		&0.2 	&0.4380* {\scriptsize(+2.3\%)}		&0.2 \\
\hline
\end{tabular}}
\end{table*}
\begin{table*}
\caption{ BM25 \& TFIDF for Disks 4\&5: train with 150 topics
and test with 100 topics. $^b$ and $^t$ denote best
and trained values respectively. * (**), $\dagger$ and $w$ as defined in Table~\ref{table:TF-IDF}.}
\centering
\label{table:cross}
\resizebox{130mm}{!}{
\begin{tabular}{|l||llll|llll|}
\hline
				&MAP$^{t}$				&MAP$^{b}$				&$w^{t}$	&$w^{b}$	&P@10$^{t}$	&P@10$^{b}$ & $w^{t}$	&$w^{b}$  \\
\hline
TFIDF baseline			&0.2398					&0.2398					&-		&-		&0.4010		&0.4010		&-	&- \\
\hline
TFIDF\_pos\_jes$\dagger$	&0.2580** {\scriptsize(+7.6\%)}		&0.2627** {\scriptsize(+9.5\%)}		&15		&10		&0.4162	{\scriptsize(+3.8\%)}	&0.4162	{\scriptsize(+3.8\%)}	&10	&10  \\
\hline
TFIDF\_pos\_ml\_wei		&\bf 0.2698** {\scriptsize(+12.5\%)}	&\bf 0.2700** {\scriptsize(+12.6\%)}	&23K		&21K		&\bf 0.4212** {\scriptsize(+5.0\%)}	&\bf 0.4222** {\scriptsize(+5.3\%)}	&22K	&20K  \\
TFIDF\_pos\_ml\_boo		&0.2450* {\scriptsize(+2.2\%)}		&0.2583** {\scriptsize(+7.7\%)}		&5K		&2K		&0.4101	{\scriptsize(+2.3\%)}	&0.4131	{\scriptsize(+3.0\%)}	&1K	&2K  \\
TFIDF\_pos\_idf			&0.2570** {\scriptsize(+7.2\%)}		&0.2639** {\scriptsize(+10.1\%)}	&5		&3		&0.4180	{\scriptsize(+4.2\%)}	&0.4202** {\scriptsize(+4.8\%)}	&3	&2  \\
TFIDF\_pos\_ridf		&0.2444* {\scriptsize(+1.9\%)}		&0.2527** {\scriptsize(+5.4\%)}		&3		&2		&0.4121	{\scriptsize(+2.8\%)}	&0.4141	{\scriptsize(+3.3\%)}	&2	&3  \\
TFIDF\_pos\_bs			&0.2456* {\scriptsize(+2.4\%)}		&0.2551** {\scriptsize(+6.4\%)}		&0.9		&0.5		&0.4131	{\scriptsize(+3.0\%)}	&0.4131	{\scriptsize(+3.0\%)}	&0.5	&0.5  \\
 \hline
\hline
BM25 baseline			&0.2621					&0.2621					&-		&-		&0.4061				&0.4061				&-	&-  \\
\hline
BM25\_pos\_jes$\dagger$		&0.2690** {\scriptsize(+2.6\%)}		&0.2690** {\scriptsize(+2.6\%)}		&2		&2		&0.4091	{\scriptsize(+0.7\%)}	&\bf 0.4172 {\scriptsize(+2.7\%)}	&3	&0.7  \\
\hline
BM25\_pos\_ml\_wei		&\bf 0.2702** {\scriptsize(+3.1\%)}	&\bf 0.2718** {\scriptsize(+3.7\%)}	&7K		&4500		&0.4111	{\scriptsize(+1.2\%)}	&0.4152	{\scriptsize(+2.2\%)}	&2K	&3K  \\
BM25\_pos\_ml\_boo		&0.2540	{\scriptsize(-3.1\%)}		&0.2671* {\scriptsize(+1.9\%)}		&1K		&200		&0.4000	{\scriptsize(-1.5\%)}	&0.4151	{\scriptsize(+2.2\%)}	&1K	&100  \\
BM25\_pos\_idf			&0.2643	{\scriptsize(+0.8\%)}		&0.2678** {\scriptsize(+2.2\%)}		&0.9		&0.4		&\bf 0.4121 {\scriptsize(+1.5\%)}	&0.4141	{\scriptsize(+2.0\%)}	&0.1	&0.2  \\
BM25\_pos\_ridf			&0.2647	{\scriptsize(+1.0\%)}		&0.2650* {\scriptsize(+1.0\%)}		&0.4		&0.2		&0.4081	{\scriptsize(+0.5\%)}	&0.4131	{\scriptsize(+1.7\%)}	&0.7	&0.2  \\  
BM25\_pos\_bs			&0.2647	{\scriptsize(+1.0\%)}		&0.2647	{\scriptsize(+1.0\%)}		&0.1		&0.1		& 0.4020 {\scriptsize(-1.0\%)}	&0.4111	{\scriptsize(+1.2\%)}	&0.2	&0.1  \\
 \hline
\end{tabular}
}
\end{table*}

Table \ref{table:cross} shows the retrieval performance of our split train-test experiments (best scores are bold). The values of the integration parameter w$^t$ and w$^b$ are those that give the best MAP and P@10 in the trainset and testset respectively. MAP$^t$ and MAP$^b$ are the MAP values obtained using w$^t$ and w$^b$ respectively in the testset (same for P@10). Similarly to Table \ref{table:TF-IDF}, all our POS-based weights outperform the baseline. In addition, the POS-based weights perform consistently to Table \ref{table:TF-IDF}, with pos\_ml\_weighted performing overall better than the rest. This consistency in the performance of the weights between Tables \ref{table:TF-IDF}-\ref{table:cross} is also seen in the $w$ values, which are very similar. These points indicate that the proposed POS-based weights are robust with respect to the values of the integration parameter $w$, for two different evaluation measures, and for different query sets. More importantly the values selected from training are portable and result in similarly good performance when testing. 
\begin{figure*}
\centering
\begin{tabular}{lr}
\resizebox{0.5\linewidth}{!}{\rotatebox{270}{\includegraphics{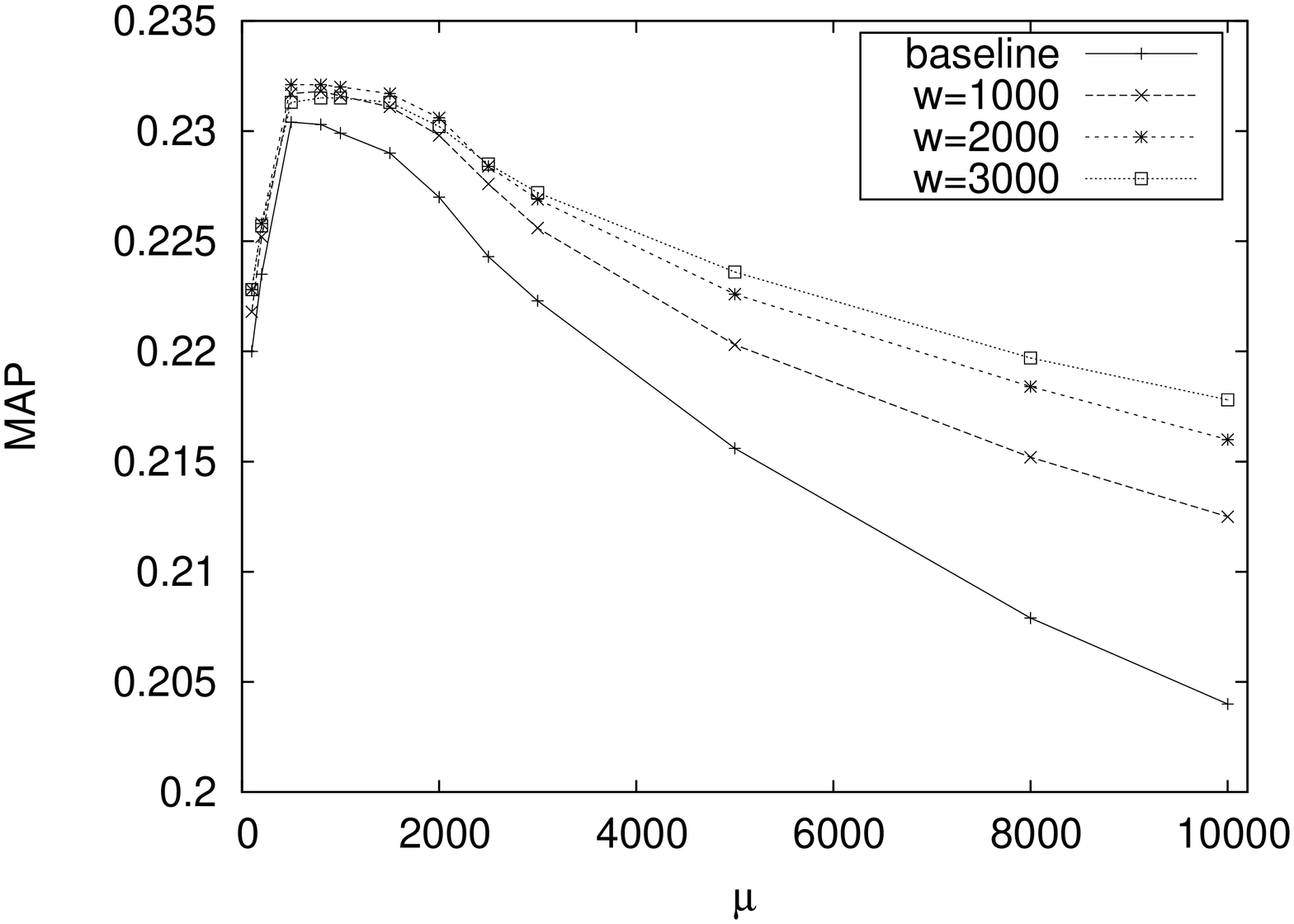}}}&
\resizebox{0.5\linewidth}{!}{\rotatebox{270}{\includegraphics{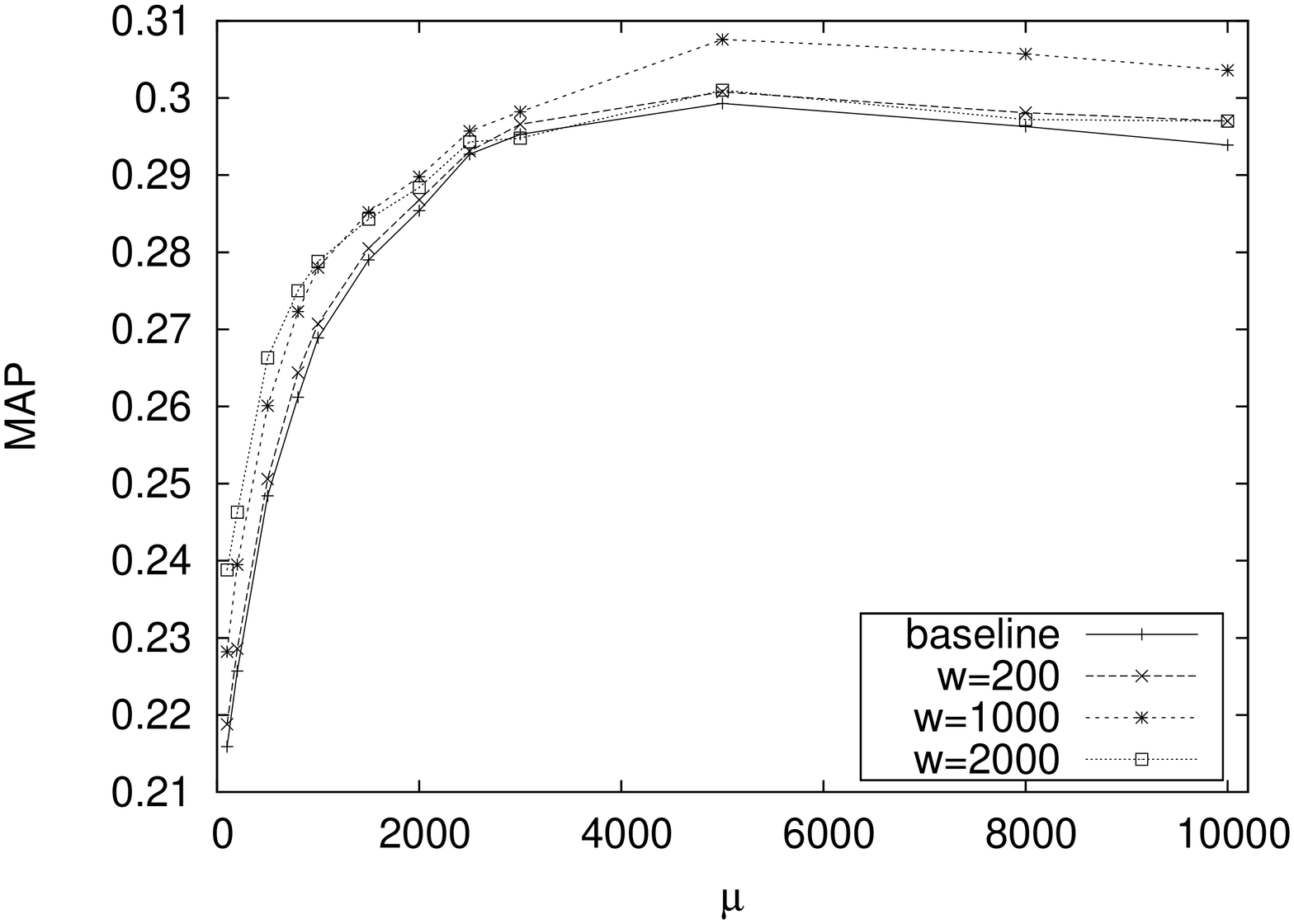}}}\\
\end{tabular}
\caption{Language Model with Dirichlet priors smoothing (baseline) \& our best POS-based term weight integrated into it ($w$ are integration parameter values).}
\label{fig:dird45}
\end{figure*}
Figure~\ref{fig:dird45} plots the MAP of the LM Dirichlet runs across the smoothing range of the retrieval model (x axis), separately for the baseline and for our best performing POS weight with three different integration values. Integrating our POS-based weight into the model always outperforms the baseline. This indicates that the contribution of our weight to retrieval is not accidental, neither due to weak tuning of the baseline, but relatively robust (for this dataset).

We compute the correlation between the proposed term weights and IDF using Spearman's $\rho$. pos\_idf is correlated with IDF ($\rho \approx 0.8$) and pos\_ridf and pos\_bs are very weakly negatively correlated with IDF. These results hold for both collections. The pos\_ml weights are not correlated with IDF. This indicates that the contribution of our POS-based term weights is different to that of IDF.

\section{Conclusion}
\label{s:Conclusion}
We proposed a new type of term weight, computed from part of speech (POS) n-grams, which represents how informative a term is in general, based on the `POS contexts' in which it generally occurs in language. We suggested five different computations of POS-based term weights by extending existing statistical approximations of term information measures. We applied these POS-based term weights to IR, by integrating them into the model that matches documents to queries. Experiments with standard TREC settings on default and tuned baselines showed that integrating our POS-based term weights to retrieval improved performance at all times. 
Future research directions include approximating our weights from more refined smoothing techniques, for instance Laplace or Good-Turing smoothing, refining the integration of our weights into retrieval, namely by treating them as prior probabilities, or applying POS-based term weighting to `flag' difficult search terms in IR. Note that these weights could also be applied to other areas, e.g., in classification, as a classification feature or threshold; in machine translation, to look at whether POS-based term weights are consistent in parallel text; and in summarisation, as an indication of general content.

\paragraph{ \textbf{Acknowledgements.}} We thank Leif Azzopardi for his valuable comments. 
Author 1 is partly funded by K.U.L. Postdoctoral Fellowship F+/08/002.
Author 2 is co-funded by FEDER, SEUI \& Xunta de Galicia under projects TIN2005-08525-C02, TIN2008-06566-C04-04/TIN \& 07SIN005206PR.

%\bibliographystyle{abbrv}
%\bibliography{mybib}

\end{document}